\documentclass[prl,aps,nofootinbib,twocolumn]{revtex4}
\usepackage{graphicx,color,amsmath,amsxtra}
\usepackage{amssymb}
\usepackage[english]{babel}
\usepackage{amsfonts}
\allowdisplaybreaks[4]

\begin{document}

\title{Unified description of dark energy and dark matter in mimetic matter model}
	
\author{Jiro Matsumoto}
\email{jmatsumoto@kpfu.ru}
\affiliation{Institute of Physics, Kazan Federal University, Kremlevskaya Street 18,
Kazan 420008, Russia}
	
\begin{abstract}
The existence of dark matter and dark energy in cosmology is implied by various observations, however, 
they are still unclear because they have not been directly detected. 
In this Letter, an unified model of dark energy and dark matter 
that can explain the evolution history of the Universe later than inflationary era, 
the time evolution of the growth rate function of the matter density contrast, 
the flat rotation curves of the spiral galaxies, and the gravitational experiments in the solar system 
is proposed in mimetic matter model. 

\end{abstract}
	
	
\maketitle

\paragraph*{Introduction.}
Dark energy and dark matter are main research themes in the current cosmology. 
Dark energy is introduced to explain the current accelerated expansion of the Universe, 
which is clarified by the observations of Type Ia supernovae \cite{Riess:1998cb,Perlmutter:1998np}, while, 
dark matter is first introduced to explain the dynamics of the galaxy clusters or 
the rotation curves of the galaxies. 
The existence of dark energy and dark matter is now strongly 
supported by the observations of Cosmic Microwave Background Radiation (CMB) \cite{Komatsu:2010fb,Ade:2013zuv,Ade:2015xua}, 
Baryon Acoustic Oscillations (BAO) \cite{Percival:2009xn,Blake:2011en,Beutler:2011hx,Cuesta:2015mqa,Delubac:2014aqe}, 
Large Scale Structure of the Universe, and so on. 
The most popular model that contain dark energy and dark matter is the $\Lambda$CDM model, 
which is composed from the cosmological constant $\Lambda$ and cold dark matter. 
The $\Lambda$CDM model is almost consistent with the observations, 
however, it is a phenomenological model; accordingly the other possibilities have been also energetically studied. 
In this Letter, we will consider mimetic matter model
\cite{Chamseddine:2013kea},
which was first proposed as a model of dark matter. However, it was realized
later that it can be treated
as a model of dark energy by adding the potential term of the scalar field
\cite{Chamseddine:2014vna}.
Mimetic matter model is a conformally invariant theory by making the physical
metric as a
product of an auxiliary metric and the contraction of an auxiliary metric and
the kinetic term of the scalar field. 
We will consider the time evolution of the Universe, 
that of the matter density perturbation, the rotation curves of the 
galaxies, and the solar system tests of gravity in a specific model of mimetic matter. 
In the following, the units of $k_\mathrm{B} = c = \hbar = 1$ are used and 
gravitational constant $8 \pi G$ is denoted by
${\kappa}^2 \equiv 8\pi/{M_{\mathrm{Pl}}}^2$ 
with the Planck mass of $M_{\mathrm{Pl}} = G^{-1/2} = 1.2 \times 10^{19}$GeV. 
\paragraph*{Time evolution of the background space-time.} 
The action of mimetic matter model we consider is given by 
\begin{align}
S=& \int d^4 x \sqrt{-g(\tilde{g}_{\mu \nu},\phi)} \bigg [ \frac{1}{2 \kappa ^2} R(g_{\mu \nu}(\tilde{g}_{\mu \nu},\phi)) 
-V(\phi) \bigg ] \nonumber \\
&+S_\mathrm{matter}, 
\label{5}
\end{align}
where $g_{\mu \nu} =\tilde{g}_{\mu \nu} \tilde{g}^{\alpha \beta} \partial _\alpha \phi \partial_\beta \phi$ \cite{Chamseddine:2014vna}. 
The action (\ref{5}) is conformally invariant with respect to the auxiliary metric $\tilde{g}_{\mu \nu}$. 
Whereas, the equivalence between mimetic matter model and
a scalar field model with a Lagrange multiplier \cite{Lim:2010yk}: 
\begin{align}
S=& \int d^4 x \sqrt{-g} \bigg [ \frac{1}{2 \kappa ^2} R(g_{\mu \nu}) -V(\phi) 
+ \lambda (g^{\mu \nu} \partial _\mu \phi \partial _\nu \phi +1) \bigg ] \nonumber \\ 
&+S_\mathrm{matter},
\label{10}
\end{align}
is shown in \cite{Golovnev:2013jxa,Barvinsky:2013mea}.
We use the representation (\ref{10}) in the following, because it simplifies the expressions of the equations. 
The potential term $V(\phi)$ is treated as a general function of $\phi$ in the 
equations which come from the principle of least action, 
while, in the concrete analyses, we consider 
the following potential function: 
\begin{equation}
V(\phi)= V_1 \mathrm{e}^{m_1^2 \phi ^2 -m_2^4 \phi ^4 } + V_2 \mathrm{e}^{-  m_3^4 \phi ^4}, 
\label{potential}
\end{equation}
where $V_1$ and $V_2$ are constants of mass dimension four, $m_1$, $m_2$, and $m_3$ are 
positive constants of mass dimension one. 
The Einstein equation obtained from Eq.~(\ref{10}) is
\begin{align}
R_{\mu \nu} - \frac{1}{2}g_{\mu \nu} R = - \kappa ^2 g_{\mu \nu}V(\phi)
-2 \kappa ^2 \lambda \partial _\mu \phi \partial _\nu \phi \nonumber \\
+ \kappa ^2 g_{\mu
\nu} \lambda (g^{\rho \sigma} \partial _\rho \phi \partial _\sigma \phi +1)
+ \kappa ^2 T_{\mu \nu},
\label{EE}
\end{align}
where $R_{\mu \nu} = \partial _\sigma \Gamma ^\sigma _{\mu \nu} - \partial _\mu
\Gamma ^\sigma _{\nu \sigma}
+ \Gamma ^\sigma _{\mu \nu} \Gamma ^\rho _{ \sigma \rho}
- \Gamma ^\sigma _{\mu \rho} \Gamma ^\rho _{\nu \sigma} $
and $T_{\mu \nu}$ is the energy momentum tensor of the usual matter.
On the other hand, the equation given by the variation with respect to the
scalar field $\phi$ is the following one:
\begin{equation}
-V_{, \phi} -2 \nabla ^\mu (\lambda \partial _\mu \phi)=0,
\label{FE}
\end{equation}
where $V_{, \phi}$ $\equiv$ $dV(\phi)/d \phi$. The constraint equation given by
the variation of $\lambda$ is
\begin{equation}
g^{\mu \nu} \partial _\mu \phi \partial _\nu \phi +1 =0.
\label{LAMBDA}
\end{equation}
When the spatially flat Friedmann-Lemaitre-Robertson-Walker (FLRW) metric,
$ds^2 =-dt^2+a^2(t)\sum_{i=1}^3 dx^i dx^i$, is taken, 
Eq.~(\ref{LAMBDA}) is expressed as 
\begin{equation}
\dot \phi ^2 = 1. 
\label{20}
\end{equation}
Then, the Friedmann equations are written by
\begin{align}
3H^2 =  \kappa ^2 \Sigma _{i=b,r} \rho _i -2 \kappa ^2 \lambda +\kappa ^2 V \label{FL00}, \\
-2 \dot H -3H^2 =   \kappa ^2 \Sigma _{i=b,r} w_i \rho _i - \kappa ^2 V 
\label{FLii},
\end{align}
where $H \equiv \dot a(t) / a(t)$.
$\rho _i$ is the energy density of the usual matter and $w_i$ is the equation of state
(EoS) parameter expressed by $w_i=p_i/ \rho _i$. The subscripts $b$ and $r$ represent 
baryon and radiation, respectively. 
From Eq.~(\ref{FE}) we have
\begin{equation}
\dot \lambda + 3H \lambda - \frac{1}{2} \dot \phi V_{,\phi} =0. \label{BFE}
\end{equation}
The equation of continuity of the matter is
\begin{equation}
\dot \rho _i +3(1+w_i) H \rho _i = 0.
\label{30}
\end{equation}
If $V_{,\phi}=0$, Eq.~(\ref{BFE}) is same as Eq.~(\ref{30}) with $w=0$. 
Therefore, $\lambda$ behaves as nonrelativistic matter. 
While, Eq.~(\ref{20}), in general, yields 
\begin{equation}
\phi (t) = \pm t +const. .
\label{40}
\end{equation}
We assume $const. =0$ in Eq.~(\ref{40}) for simplicity in this Letter. 

Let us consider the concrete form of the potential function (\ref{potential}). 
If $V_1 \sim M_\mathrm{pl}^2H_0^2$, $m_1 \sim m_2 \sim H_0$, 
$V_2 \sim 10^5 M_\mathrm{pl}^2H_0^2$, and $m_3 \sim 10^5 H_0$, 
where $H_0$ means that the Hubble constant in the $\Lambda$CDM model $H_0 \simeq 68$ (km/s)/Mpc, are assumed, 
then the term proportional to 
$V_1$ behaves as dark energy and the term proportional to 
$V_2$ is not contributed to the evolution of the Universe, 
because the energy density of nonrelativistic matter at $t \sim m_3 ^{-1}$: $\rho _\mathrm{m} \sim 10^{14}M_\mathrm{pl}^2H_0^2$, is 
much more than $V_2$, moreover, the suppression term $\mathrm{e}^{-m_3^4 \phi ^4}$ is at work when $t > m_3 ^{-1}$. 
The reason why the values of $V_2$ and $m_3$ are assumed as above is to explain 
the flat rotation curves of the spiral galaxies as shown in the following paragraph. 
By the way, $V(\phi)$ is almost constant in the regime $t < 10^{-1}H_0^{-1}$, then 
$\lambda (t)$ is approximately expressed as $\lambda (t) = \lambda _0 a^{-3}(t)$ by Eq.~(\ref{BFE}). 
If the constant $\lambda _0$ is set to be $\lambda _0 = - \rho _{\mathrm{DM},0}/2$, 
where $\rho _{\mathrm{DM},0}$ is the current energy density of dark matter in the $\Lambda$CDM model, 
then the term $-2 \kappa ^2 \lambda$ in Eq.~(\ref{FL00}) behaves as dark matter. 
We use $\lambda _0 = - \rho _{\mathrm{DM},0}/2$ to realize the same expansion history 
of the early universe as that in the $\Lambda$CDM model in this Letter. 

\begin{figure}[h]
\begin{center}
\includegraphics[clip, width=0.97\columnwidth]{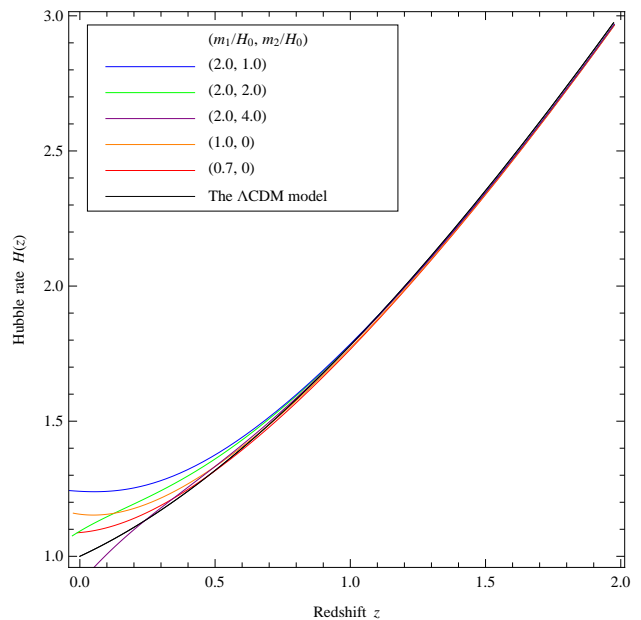}
\end{center}
\caption{Redshift dependence of the Hubble rate function. The 
Hubble rate function is normalized by the Hubble constant of the $\Lambda$CDM model ``$H_0$".  
$\Omega _\mathrm{m}=0.31$, $\Omega _\Lambda=0.69$ is assumed in the $\Lambda$CDM model. 
While, $V_1$ is fixed by $V_1=1.8 \kappa ^2 H_0^2$. }
\label{f1}
\end{figure}

Figure~\ref{f1} shows the comparisons of the Hubble rate function between our model and the $\Lambda$CDM model. 
The case that the Hubble rate function is larger than that of the $\Lambda$CDM model is in particular 
depicted in Fig.~\ref{f1}, because the recent observations of Type Ia supernovae 
revealed such a behavior in the low-redshift region \cite{Riess:2016jrr,Riess:2011yx,DiValentino:2016hlg,Marra:2013rba}. 
Of course, our model can be consistent with the $\Lambda$CDM model if $m_1=m_2=0$. 

\begin{figure}[h]
\begin{center}
\includegraphics[clip, width=0.97\columnwidth]{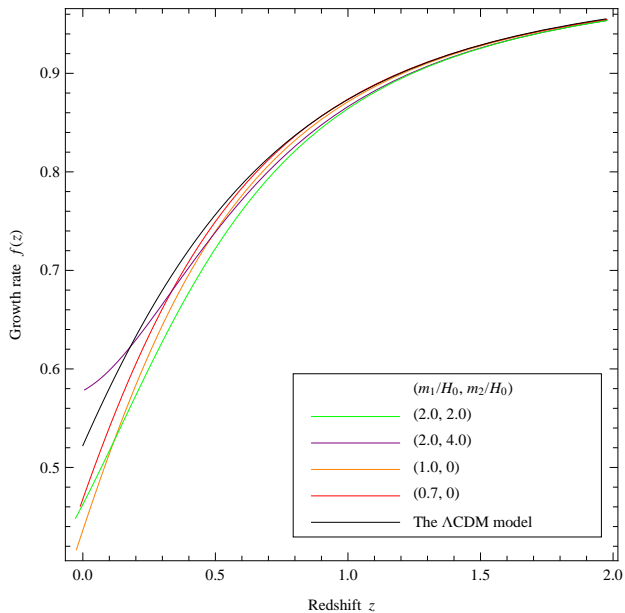}
\end{center}
\caption{Evolution of the growth rate function $f(z)= d \ln \delta /d N$. The initial conditions are 
assigned as $f''(10)=f'(10)=0$, and $f(10)=1$. The values of the parameters are same as those in Fig.~\ref{f1}. }
\label{f2}
\end{figure}

\paragraph*{Matter density perturbation.}
The master equation of the matter density contrast $\delta \equiv \delta \rho _b/ \rho _b$ 
in mimetic matter model is given by \cite{Matsumoto:2015wja}
\begin{align}
\ddddot \delta + \left ( 7H- \frac{\dot \lambda}{\lambda} \right ) \dddot \delta
+ \left ( 16H^2 -4 \kappa ^2 (\rho _b- 2 \lambda) -5H \frac{\dot \lambda}{\lambda} \right ) \ddot \delta 
\nonumber \\
+ \frac{3}{2} \left ( 8H^3 - 3\kappa ^2 H(\rho _b- 4 \lambda) 
-4H^2 \frac{\dot \lambda}{\lambda} + \kappa ^2 \rho _b \frac{\dot \lambda}{\lambda} \right ) \dot \delta =0. 
\label{50}
\end{align}
Equation (\ref{50}) looks complicated, however, 
it shows that a quite similar behavior of the matter density contrast 
to that in the $\Lambda$CDM model. 
In the case of $V_\phi =0$, the master equation of the matter density contrast can be written as \cite{Matsumoto:2015wja}
\begin{equation}
\ddot \delta _\mathrm{tot} + 2H \dot \delta _\mathrm{tot} - \frac{\kappa ^2}{2} \rho _\mathrm{tot} \delta _\mathrm{tot}=0, 
\label{60}
\end{equation}
where $\rho _\mathrm{tot} \equiv \rho _b -2 \lambda$, 
$\delta _\mathrm{tot} \equiv (\delta \rho _b -2 \delta \lambda)/(\rho _b -2 \lambda)$. 
If we regard $\rho _\mathrm{tot}$ as $\rho _\mathrm{m}=\rho _b + \rho _\mathrm{DM}$, 
Eq.~(\ref{60}) is completely equivalent with the equation of the matter density contrast in the $\Lambda$CDM model. 
Paying attention to the fact that $V_{, \phi} \simeq 0$ is held around $z = 10$ enables 
us to use $f(z=10) \simeq 1$, where $f \equiv d \ln \delta /d N$ is the growth factor of the matter density contrast 
and $N= \ln a$, as an initial condition of the 
numerical calculations. 
The calculation results of the growth factor are shown in Fig.~\ref{f2}. 
The deviations from the $\Lambda$CDM model in Fig.~\ref{f2} appear in higher redshift and they 
become larger compared to those in Fig.~\ref{f1}. 
While, we can recognize that the lower growth rate is realized when the Hubble rate is higher as expected. 
\paragraph*{Spherically symmetric solutions.}
In the static space-time with spherical symmetry, 
$ds^2 = -\mathrm{e}^{2 \Phi (r)}dt^2+\mathrm{e}^{2 \Psi (r)} dr^2 +r^2(d \theta ^2 + \sin ^2 \theta d \varphi ^2)$, 
the constraint equation (\ref{LAMBDA}) is written by
\begin{equation}
\mathrm{e}^{-2 \Psi} \phi ^{\prime 2} +1 =0 \label{sp4}. 
\end{equation}
Then, the Einstein equations in the case $\rho = p  =0$ are given as 
\begin{align}
\frac{1}{\kappa ^2} \left ( \frac{1}{r^2} - \frac{2 \Psi '}{r} \right ) \mathrm{e} ^{-2 \Psi} - \frac{1}{\kappa ^2 r^2} 
= -V, \label{sp6} \\
\frac{1}{\kappa ^2} \left ( \frac{1}{r^2} + \frac{2 \Phi '}{r} \right ) \mathrm{e} ^{-2 \Psi} - \frac{1}{\kappa ^2 r^2} 
= -V+2 \lambda, \label{sp7} \\
\frac{1}{\kappa ^2} \left ( \Phi '' + \Phi ^{\prime 2} - \Psi ' \Phi ' + \frac{\Phi '}{r} - \frac{\Psi '}{r} \right ) 
\mathrm{e} ^{-2 \Psi} = -V.  \label{sp8}
\end{align}
The equation of the scalar field (\ref{FE}) yields
\begin{equation}
\lambda ' + \left ( \Phi ' +\frac{2}{r} \right ) \lambda - \frac{1}{2} \phi ' V_{, \phi} = 0. \label{sp9}
\end{equation}
If $V_\phi =0$, it is possible to analyze the above equations, because Eq.~(\ref{sp9}) can be exactly solved 
\cite{Deruelle:2014zza,Myrzakulov:2015sea}. 
It is, in general, difficult to solve Eqs.~(\ref{sp6})-(\ref{sp9}), 
because of the nonlinearity of the equations. 
However, if we assume the conditions $\vert \Psi \vert , \vert r \Psi ' \vert,  \vert \Phi \vert , \vert r \Phi ' \vert \ll 1$, 
we can evaluate the behavior of the functions $\Psi (r)$ and $\Phi (r)$ as follows. 
First, Eq.~(\ref{sp4}) gives 
\begin{equation}
\phi (r) = \pm i r + const. 
\label{sp10}
\end{equation}
If we fix $const. =0$ in Eq.~(\ref{sp10}), the field $\phi (r)$ becomes pure imaginary. 
One may think it is problematic. 
However, there is no problem as long as the potential $V(\phi)$ is an even function, 
because the equations are, then, only described by the real functions. 
The concrete form of $\lambda (r)$ is 
given by solving Eq.~(\ref{sp9}) in the limit $\vert r \Phi ' \vert \ll 1$: 
\begin{equation}
\lambda '  +\frac{2}{r} \lambda - \frac{1}{2} \phi ' V_{, \phi} = 0. 
\label{sp13}
\end{equation}
While, if we express $\Phi (r)$ and $\Psi (r)$ as 
$\Phi (r) = r_0 / r + \delta \Phi (r)$, where $r_0$ is a constant, and $\Psi (r)= - \Phi (r) + \delta \Psi (r)$, 
then Eqs.~(\ref{sp6}) and (\ref{sp7}) yield
\begin{align}
\delta \Psi ' (r) = \kappa ^2 r \lambda (r), \label{sp11} \\
\partial _r (r \delta \Phi (r)) = \delta \Psi (r) - \frac{\kappa ^2 r^2}{2} V(r) + \kappa ^2 r^2 \lambda (r). \label{sp12}
\end{align}
Here, $\delta \Phi (r)$ and $\delta \Psi (r)$ express the deviations from the 
vacuum solution $\Phi (r) = - \Psi (r) = r_0/r$. 
Equations (\ref{sp11}) and (\ref{sp12}) give the explicit forms of 
$\Phi (r)$ and $\Psi (r)$, however, we should remember that we first assume the conditions 
$\vert \Psi \vert , \vert r \Psi ' \vert,  \vert \Phi \vert , \vert r \Phi ' \vert \ll 1$. 
By taking into account that $\vert r_0/r \vert = GM/r \ll 1$ is usually 
satisfied (e.g. $GM_\odot/R_\odot \sim 10^{-6}$), what we should make sure is the conditions 
$\vert \delta \Psi \vert , \vert r \delta \Psi ' \vert,  \vert \delta \Phi \vert , \vert r \delta \Phi ' \vert \ll 1$. 
Substitutions of Eqs.~(\ref{sp11}) and (\ref{sp12}) into 
$\vert \delta \Psi \vert , \vert r \delta \Psi ' \vert,  \vert \delta \Phi \vert , \vert r \delta \Phi ' \vert \ll 1$, 
in general, give a constraint for $r$. In other words, the conditions 
$\vert \delta \Psi \vert , \vert r \delta \Psi ' \vert,  \vert \delta \Phi \vert , \vert r \delta \Phi ' \vert \ll 1$ 
give a domain of the solutions (\ref{sp11}) and (\ref{sp12}). 
Then, the numerical calculations can be executed by regarding the solutions (\ref{sp11}) and (\ref{sp12}) as the boundary conditions, 
and the behavior of the solutions in the whole region will be clarified. 

Let us consider the case that potential function is given by Eq.~(\ref{potential}). 
In the region that $m_1 r \sim m_2 r \ll 1$, and $m_3 r \ll 1$ are satisfied, 
the potential function $V(\phi)$ and its derivative are approximately expressed as $V(r)\sim V_1(1  -m_1^2r^2) +V_2 (1-m_3^4r^4)$, 
and $\phi 'V_{, \phi} (r) \sim -2(V_1m_1^2 +2V_2 m_3^4r^2)r$. Then, Eq.~(\ref{sp13}) gives 
\begin{equation}
\lambda (r)= \frac{s_1}{r^2} - \frac{1 }{4}V_1m_1^2 r^2 - \frac{1}{3}V_2m_3^4r^4, 
\label{sp14}
\end{equation}
where $s_1$ is an arbitrary constant. 
Substituting Eq.~(\ref{sp14}) into Eqs.~(\ref{sp11}) and (\ref{sp12}) yields
\begin{align}
\delta \Psi (r) = c_1 + \kappa ^2 s_1 \ln \frac{r}{r_1} - \frac{1}{16} \kappa ^2 V_1m_1^2 r^4 \nonumber \\
- \frac{1}{18}\kappa ^2 V_2 m_3^4 r^6, \label{sp15} \\
\delta \Phi (r) = \frac{r_2}{r} + c_1 + \kappa ^2 s_1 \ln \frac{r}{r_1} - \frac{\kappa ^2}{6} (V_1 +V_2)r^2 \nonumber \\
+ \frac{3}{80}\kappa ^2 V_1 m_1^2r^4 + \frac{1}{63}\kappa ^2 V_2 m_3^4 r^6, \label{sp16}
\end{align}
where $r_1$, $r_2$, and $c_1$ are arbitrary constants. 
When the arbitrary constants $s_1$, $r_2$, and $c_1$ are negligibly small, 
$\vert \delta \Psi \vert \ll \vert \delta \Phi \vert \sim \kappa ^2 \vert V_1 +V_2 \vert r^2 /6$ 
are held. 
Then, the condition $\vert \delta \Phi \vert \ll 1$ gives a constraint for $r$: $r \ll 10^{-2}H_0^{-1}$. 
Here, $V_1 \sim H_0^2/\kappa ^2$ and $V_2 \sim 10^5 H_0^2/\kappa ^2$ are assumed. 
On the other hand, the conditions $m_1 r \sim m_2 r \ll 1$ and $m_3 r \ll 1$ 
give $r \ll 10^{-5}H_0^{-1}$ if $m_1 \sim m_2 \sim H_0$, $m_3 \sim 10^5 H_0$. 
Therefore, Eqs.~(\ref{sp14})-(\ref{sp16}) are valid in the region $r \ll 10^{-5}H_0^{-1}$. 

\begin{figure}[h]
\begin{center}
\includegraphics[clip, width=0.97\columnwidth]{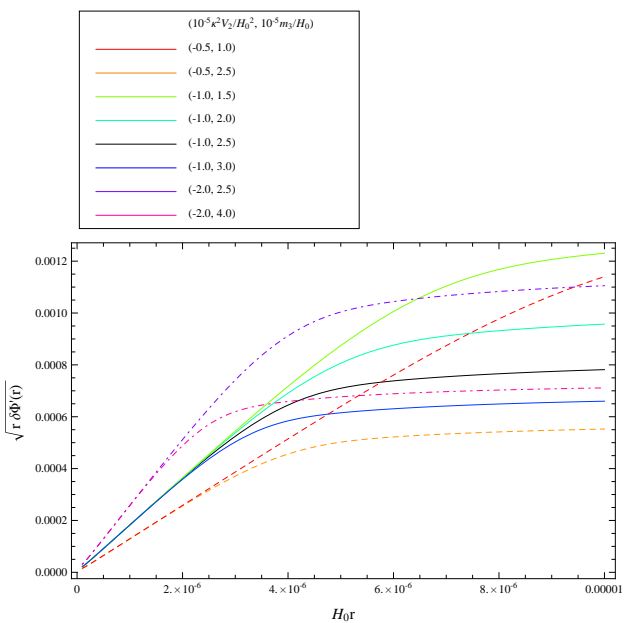}
\end{center}
\caption{Contributions to the rotational velocity from mimetic matter. 
Here, the scale $10^{-6}cH_0^{-1}$ corresponds to $\sim 4$ kpc, 
and $v \simeq 3 \sqrt{r \delta \Phi ' (r)} \times 10^5$ km s$^{-1}$ is given when 
$\Phi '(r) \simeq \delta \Phi '(r)$. The arbitrary constants in Eqs.~(\ref{sp14})-(\ref{sp16}) are fixed as $s_1$, $r_2$, $c_1=0$.}
\label{f3}
\end{figure}

The behavior of the solutions around $r = 10^{-5}H_0^{-1}$ are 
also comprehended by using Eqs.~(\ref{sp14})-(\ref{sp16}) as boundary conditions. 
The radius dependence of $\sqrt{r \delta \Phi '(r)}$ in the region $r=10^{-6}H_0^{-1}-10^{-5}H_0^{-1}$ 
is shown in Fig.~\ref{f3}. 
$\delta \Phi (r)$ is well approximated as $\delta \Phi (r) \simeq - \kappa ^2 V_2 r^2/6$ around $r=10^{-6}H_0^{-1}$, 
while, the damping factor $\mathrm{e}^{-  m_3^4 \phi ^4}$ in Eq.~(\ref{potential}) 
becomes effective when $r$ is larger than $10^{-6}H_0^{-1}$, 
therefore, the inclination of $\sqrt{r \delta \Phi '(r)}$ becomes gentler than a straight line as seen in Fig.~\ref{f3}. 
The reason why $\sqrt{r \delta \Phi '(r)}$ is plotted is that 
the rotation speed of a galaxy is approximately given by 
$v=\sqrt{r \Phi ' (r)}=\sqrt{r\Phi _0'(r) + r \delta \Phi '(r)}$, where $\Phi _0 (r)$ is the 
Newtonian potential of the galaxy, if 
the circular motion is assumed. 
Namely, $\sqrt{r \delta \Phi '(r)}$ represents the corrections for the rotation speed from mimetic matter. 
The curves in Fig.~\ref{f3} may look rather steep, however, the potential made by baryons in the galaxy: $r\Phi _0'(r)$ 
will make the rotation curve flatter. 

The gravitational behavior of our model in the solar system is given by 
\begin{align}
\delta \Psi (r) = c_1 + \kappa ^2 s_1 \ln \frac{r}{r_1} , \label{sp17} \\
\delta \Phi (r) = \frac{r_2}{r} + c_1 + \kappa ^2 s_1 \ln \frac{r}{r_1} - \frac{\kappa ^2}{6} (V_1 +V_2)r^2 , \label{sp18}
\end{align}
because the radius $r$ is enough small. 
If the arbitrary constants are fixed as $c_1=s_1=r_2=0$, then 
the solutions (\ref{sp17}) and (\ref{sp18}) are consistent with the 
Schwarzschild-de Sitter solution. 
We should note that $V_2$ is greater than usual cosmological constant by five orders of magnitude. 
However, it is known that the metric functions in Schwarzschild-de Sitter space-time can pass the solar system tests 
even if the value of the cosmological constant is greater than usual one 
by five orders of magnitude \cite{Kagramanova:2006ax}. 
\paragraph*{Conclusions.}
In this Letter, an example of the unified discription of 
dark energy and dark matter in mimetic matter model has been shown. 
This model can also describe the same time evolution of the Universe and the matter density perturbation as those in the $\Lambda$CDM 
model, 
however, the case that the Hubble rate function is greater than that of the $\Lambda$CDM model in low-redshift region 
have been, in particular, considered, because the recent observations of supernovae show such a behavior in the Hubble rate function. 
Then, the growth rate function of the matter density perturbation 
becomes less than that of the $\Lambda$CDM model. It is also consistent with the observational results. 
Whereas, the term $V_2 \mathrm{e}^{-m_3^4 \phi ^4}$ 
has been introduced to explain the flat rotation curves of the spiral galaxies. 
This term does not influence the background evolution of the Universe, but influence the 
galaxy-scale physics. Moreover, it can easily pass the solar system tests of gravity. 
While, the corrections for gravity in the galaxy scale are independent from the mass of the galaxies. 
They only depend on the parameters $V_2$ and $m_3$. 
Therefore, detailed studies of the baryon density distribution of the galaxies will clarify whether or not this model is valid. 
\paragraph*{Appendix.}
The work was supported by the Russian Government Program of Competitive Growth of Kazan Federal University. 


\begin{thebibliography}{99}
\bibitem{Riess:1998cb} 
  A.~G.~Riess {\it et al.} [Supernova Search Team Collaboration],
  Astron.\ J.\  {\bf 116}, 1009 (1998)
  [astro-ph/9805201].

\bibitem{Perlmutter:1998np} 
  S.~Perlmutter {\it et al.} [Supernova Cosmology Project Collaboration],
  Astrophys.\ J.\  {\bf 517}, 565 (1999)
  [astro-ph/9812133]. 
  
\bibitem{Komatsu:2010fb} 
  E.~Komatsu {\it et al.} [WMAP Collaboration],
  Astrophys.\ J.\ Suppl.\  {\bf 192}, 18 (2011)
  [arXiv:1001.4538 [astro-ph.CO]].

\bibitem{Ade:2013zuv} 
  P.~A.~R.~Ade {\it et al.} [Planck Collaboration],
  Astron.\ Astrophys.\  {\bf 571}, A16 (2014)
  [arXiv:1303.5076 [astro-ph.CO]].

\bibitem{Ade:2015xua} 
  P.~A.~R.~Ade {\it et al.} [Planck Collaboration],
  arXiv:1502.01589 [astro-ph.CO].
  
\bibitem{Percival:2009xn} 
  W.~J.~Percival {\it et al.} [SDSS Collaboration],
  Mon.\ Not.\ Roy.\ Astron.\ Soc.\  {\bf 401}, 2148 (2010)
  [arXiv:0907.1660 [astro-ph.CO]].

\bibitem{Blake:2011en} 
  C.~Blake {\it et al.},
  Mon.\ Not.\ Roy.\ Astron.\ Soc.\  {\bf 418}, 1707 (2011)
  [arXiv:1108.2635 [astro-ph.CO]].

\bibitem{Beutler:2011hx} 
  F.~Beutler {\it et al.},
  Mon.\ Not.\ Roy.\ Astron.\ Soc.\  {\bf 416}, 3017 (2011)
  [arXiv:1106.3366 [astro-ph.CO]].

\bibitem{Cuesta:2015mqa} 
  A.~J.~Cuesta {\it et al.},
  Mon.\ Not.\ Roy.\ Astron.\ Soc.\  {\bf 457}, no. 2, 1770 (2016)
  [arXiv:1509.06371 [astro-ph.CO]].

\bibitem{Delubac:2014aqe} 
  T.~Delubac {\it et al.} [BOSS Collaboration],
  Astron.\ Astrophys.\  {\bf 574}, A59 (2015)
  [arXiv:1404.1801 [astro-ph.CO]].  

\bibitem{Chamseddine:2013kea}
    A.~H.~Chamseddine and V.~Mukhanov,
    JHEP {\bf 1311}, 135 (2013)
    [arXiv:1308.5410 [astro-ph.CO]].

\bibitem{Chamseddine:2014vna}
    A.~H.~Chamseddine, V.~Mukhanov and A.~Vikman,
    JCAP {\bf 1406}, 017 (2014)
    [arXiv:1403.3961 [astro-ph.CO]].

\bibitem{Lim:2010yk} 
  E.~A.~Lim, I.~Sawicki and A.~Vikman,
  JCAP {\bf 1005}, 012 (2010)
  [arXiv:1003.5751 [astro-ph.CO]].

\bibitem{Golovnev:2013jxa}
    A.~Golovnev,
    Phys.\ Lett.\ B {\bf 728}, 39 (2014)
    [arXiv:1310.2790 [gr-qc]].

\bibitem{Barvinsky:2013mea}
    A.~O.~Barvinsky,
    JCAP {\bf 1401}, no. 01, 014 (2014)
    [arXiv:1311.3111 [hep-th]]. 

\bibitem{Riess:2016jrr}
A.~G.~Riess {\it et al.},
Astrophys.\ J.\ {\bf 826}, no. 1, 56 (2016)
[arXiv:1604.01424 [astro-ph.CO]].

\bibitem{Riess:2011yx} 
  A.~G.~Riess {\it et al.},
  Astrophys.\ J.\  {\bf 730}, 119 (2011)
  Erratum: [Astrophys.\ J.\  {\bf 732}, 129 (2011)]
  [arXiv:1103.2976 [astro-ph.CO]].

\bibitem{DiValentino:2016hlg} 
  E.~Di Valentino, A.~Melchiorri and J.~Silk,
  Phys.\ Lett.\ B {\bf 761}, 242 (2016)
  [arXiv:1606.00634 [astro-ph.CO]].

\bibitem{Marra:2013rba} 
  V.~Marra, L.~Amendola, I.~Sawicki and W.~Valkenburg,
  Phys.\ Rev.\ Lett.\  {\bf 110}, no. 24, 241305 (2013)
  [arXiv:1303.3121 [astro-ph.CO]].

\bibitem{Matsumoto:2015wja} 
  J.~Matsumoto, S.~D.~Odintsov and S.~V.~Sushkov,
  Phys.\ Rev.\ D {\bf 91}, no. 6, 064062 (2015)
  [arXiv:1501.02149 [gr-qc]].

\bibitem{Deruelle:2014zza} 
  N.~Deruelle and J.~Rua,
  JCAP {\bf 1409}, 002 (2014)
  [arXiv:1407.0825 [gr-qc]].

\bibitem{Myrzakulov:2015sea} 
  R.~Myrzakulov and L.~Sebastiani,
  Gen.\ Rel.\ Grav.\  {\bf 47}, no. 8, 89 (2015)
  [arXiv:1503.04293 [gr-qc]]. 
  
\bibitem{Kagramanova:2006ax} 
  V.~Kagramanova, J.~Kunz and C.~Lammerzahl,
  Phys.\ Lett.\ B {\bf 634}, 465 (2006)
  [gr-qc/0602002].
  
\end{thebibliography}
\end{document}